\documentclass[11pt,a4paper]{article}

\usepackage{amsmath,amsthm,amsfonts,amssymb,color,hyperref,anysize,enumitem,graphicx,epstopdf}
\usepackage[usenames,dvipsnames]{xcolor}
\usepackage[margin=0.84in]{geometry}

\newcommand{\nn}{\mathbb{N}}

\renewcommand{\O}{\mathcal{O}}

\newcommand{\la}{\leftarrow}

\newcommand{\floor}[1]{\left\lfloor #1 \right\rfloor}
\newcommand{\ceil}[1]{\left\lceil #1 \right\rceil}

\newcommand{\prob}[1]{\mathbb{P}\left[ #1 \right]}

\newcommand{\expect}[1]{\mathbb{E}\left[ #1 \right]}

\renewcommand{\bold}[1]{\textbf{#1}}
\newcommand{\parabold}[1]{\noindent\bold{#1}}

\newenvironment{pf}{\begin{proof}[\emph{\textbf{Proof: }}]}{\end{proof}}
\newenvironment{pfof}[1]{\begin{proof}[\emph{\textbf{Proof of #1: }}]}{\end{proof}}

\usepackage{tcolorbox}
\tcbuselibrary{theorems}

\newtcbtheorem[auto counter]{coloredDEF}{Definition}
{colback=blue!5,colframe=blue!40!black,fonttitle=\bfseries,before={\vspace{0.3cm}},after={\vspace{0.3cm}}}{de}

\newtcbtheorem[auto counter]{coloredNOT}{Notation}
{colback=blue!5,colframe=blue!40!black,fonttitle=\bfseries,before={\vspace{0.3cm}},after={\vspace{0.3cm}}}{no}

\newtcbtheorem[auto counter]{assume}{Assumption}
{colback=green!5,colframe=green!60!black!80,fonttitle=\bfseries,before={\vspace{0.3cm}},after={\vspace{0.3cm}}}{assume}

\newtcbtheorem[auto counter]{coloredLEM}{Lemma}
{colback=red!5,colframe=red!60!black,fonttitle=\bfseries,before={\vspace{0.3cm}},after={\vspace{0.3cm}}}{lm}

\newtcbtheorem[auto counter]{coloredTHM}{Theorem}
{colback=purple!5,colframe=purple!50!black,fonttitle=\bfseries,before={\vspace{0.3cm}},after={\vspace{0.3cm}}}{th}

\newtheorem{theorem}{Theorem}
\newtheorem{lemma}[theorem]{Lemma}

\newcommand{\calE}{\mathcal{E}}

\newcommand{\calQ}{\mathcal{Q}}

\usepackage[ruled,vlined,linesnumbered]{algorithm2e}
\usepackage{algorithmicx}

\newcommand{\Exp}{\mathsf{Exp}}

\newcommand{\dist}{\mathsf{dist}}
\newcommand{\SP}{\mathsf{SP}}
\newcommand{\TD}{\text{\textsf{Term-Detour}}}
\newcommand{\detour}{\text{\textsf{Detour}}}
\newcommand{\polyk}{\mathsf{poly}(k)}
\newcommand{\Gs}{G_*}
\newcommand{\calEearly}{\mathcal{E}^{\mathsf{early}}}
\newcommand{\calEmany}{\mathcal{E}^{\mathsf{many}}}
\newcommand{\calEfar}{\mathcal{E}^{\mathsf{far}}}
\newcommand{\erlang}{\text{\textsf{Erlang}}}
\title{Steiner Point Removal\\
--- Distant Terminals Don't (Really) Bother}
\author{Yun Kuen Cheung\\Max Planck Institute for Informatics\\Saarland Informatics Campus}
\date{}

\begin{document}
\maketitle
\begin{abstract}
Given a weighted graph $G=(V,E,w)$ with a set of $k$ terminals $T\subset V$,
the Steiner Point Removal problem seeks for a minor of the graph with vertex set $T$,
such that the distance between every pair of terminals is preserved within a small multiplicative distortion.
Kamma, Krauthgamer and Nguyen (SODA 2014, SICOMP 2015) used a ball-growing algorithm to show that
the distortion is at most $\O(\log^5 k)$ for general graphs.

In this paper, we improve the distortion bound to $\O(\log^2 k)$. The improvement is achieved based on 
a known algorithm that constructs terminal-distance exact-preservation minor with $\O(k^4)$ (which is independent of $|V|$) vertices,
and also two tail bounds on the sum of independent exponential random variables,
which allow us to show that it is unlikely for a non-terminal being contracted to a distant terminal.

\medskip

\noindent\textbf{Keywords.}~Steiner Point Removal, Graph Sparsification, Vertex Sparsification, Exponential Random Variables
\end{abstract}

\section{Introduction}
\emph{Graph Compression/Sparsification} generally describes a transformation of a \emph{large} graph into a \emph{smaller} graph
that preserves, either exactly or approximately, certain features (e.g., distance, cut, flow) of the large graph.
Its algorithmic value is clear, since the compressed graph can be computed in a preprocessing step of an algorithm,
so as to reduce subsequent running time and memory requirement.
Some notable examples are graph spanners, low-stretch spanning tree, distance oracles and cut/flow/spectral sparsifiers.

In this paper, we study a vertex sparsification problem called the \emph{Steiner Point Removal (SPR) problem}.
Given a weighted graph $G=(V,E,w)$ with a set of $k$ terminals $T\subset V$,
the SPR problem seeks for a new graph $G'$ which is a minor of $G$,
such that the vertex set of $G'$ is $T$ (i.e., no non-terminal in the minor)
and the distance between every pair of terminals is preserved within a multiplicative distortion $\kappa$:
$$
\forall t_i,t_j\in T,~~~~\dist_G(t_i,t_j) ~\le~ \dist_{G'}(t_i,t_j) ~\le~ \kappa\cdot \dist_G(t_i,t_j).
$$
The target is to minimize $\kappa$.
The requirement that $G'$ is a minor of $G$ is crucial
since minor operations preserve certain structural similarities, e.g., planarity, of the input graph $G$.

SPR problem was first initiated by Gupta~\cite{Gupta2001}, where the input graphs are trees.
He showed that for tree graphs, the distortion is at most $8$. A matching lower bound was shown by Chan et al.~\cite{CXKR2006}.
Recently, Kamma, Krauthgamer and Nguyen\footnote{Since we will refer to this paper for multiple times,
for brevity we will refer it as KKN.}~\cite{KKN2015} used a randomized ball-growing algorithm
to show the interesting result that the distortion is at most $\O(\log^5 k)$ for general graphs.
It remains a big gap between their upper bound and the best known lower bound (which is $8$);
narrowing this gap is the motivation of the current work.

In this paper, we improve the distortion bound to $\O(\log^2 k)$.
Our algorithm starts with a preprocessing step, which uses an algorithm of Krauthgamer, Nguyen and Zondiner~\cite{KNZ2014}
to obtain a minor of the input graph such that all terminal distances are preserved exactly, while the minor contains only $\O(k^4)$ vertices.
Then we use the randomized ball-growing algorithm of KKN (with a few adjustments on parameters) on the preprocessed graph.
By working with the preprocessed graph that contains only $\polyk$ vertices,
we can define a class of $\polyk$ ``bad events'' such that avoidances of all these bad events
will lead to a minor with $\O(\log^2 k)$ distortion.
Finally, we show that the algorithm of KKN avoids all bad events with high probability.

In KKN's analysis, they first proved the result for graphs with bounded diameters;
for graphs with any diameters, they needed to provide a separate argument to reduce to low-diameter scenarios.
In contrast, our analysis will work directly --- we do not need any assumption on the graph diameters.
Also, our analysis bypasses KKN's need for analyzing \emph{long active subpaths}.
Hence, our analysis is more compact than KKN's analysis.

We note that an important component of our analysis is two tail bounds on the sum of independent exponential random variables.
This allows us to show that it is unlikely for a non-terminal being contracted to a \emph{distant} terminal.

\medskip

\parabold{Further Related Work.}
Basu and Gupta~\cite{BG2008} showed that for outer-planar graphs
(planar graphs with all terminals on the outermost face), SPR problem can be solved with distortion $\O(1)$.
When randomization is allowed, Englert et al.~\cite{EnglertGKRTT2014} showed that for graphs that exclude a fixed minor,
one can construct a randomized minor for SPR problem with $\O(1)$ \emph{expected} distortion.
It remains open on whether similar guarantees can be obtained in the deterministic setting.

A natural generalization of SPR problem is to allow the minor keeping a small number of non-terminals.
This generalization was initiated by Krauthgamer, Nguyen and Zondiner~\cite{KNZ2014},
in which they focused on preserving terminal distances \emph{exactly} for general graphs, trees, planar graphs and graphs with bounded treewidth.
Cheung, Gramoz and Henzinger~\cite{CGH2016} extended to the setting of preserving terminal distances \emph{approximately}.
They proved super-linear (in $k$) lower bounds on the number of non-terminals required in the minor for achieving distortion less than $8$.
They also extended the technique for proving the lower bounds to establish the following result related to SPR problem:
if the tight bound of SPR problem is super-constant (in $k$), then allowing the freedom of keeping $\O(k)$ non-terminals in the minor
would \emph{not} improve the distortion bound to a constant.

In many other settings, problems that concern preserving features which involve only a small number of terminals are becoming more popular.
Terminal distance oracles were studied by Roditty, Thorup and Zwick~\cite{RTZ2005}. They showed that there exists
oracle using only $\O(\tau\cdot |V|\cdot k^{1/\tau})$ space which answers the distance
between a terminal and any other vertex with stretch at most $(2\tau-1)$.
Elkin, Filtser and Neiman~\cite{EFN2015-stoc} improved the space requirement to $\O(\tau\cdot k^{1/\tau} + |V|)$,
with a slightly worse stretch of at most $(4\tau-1)$.
Recently, Elkin, Filtser and Neiman~\cite{EFN2015-approx} considered the problem of embedding a finite metric (e.g., graph metric)
into normed metric while preserving all distances from any terminal to all other vertices approximately.
They showed that the distortion depends only on $k$, but not on $|V|$.
For cut/flow sparsifiers, we refer readers to~\cite{LM2010,Moitra2009,CLLM2010,MM2016,Chuzhoy2012,KR2013,AGK2014} for more details.

\section{Main Theorem and Preliminary}

For any graph $G$, $G'$ is a minor of $G$ if $G'$ can be attained from $G$
by a sequence of edge contractions, edge deletions and vertex deletions.

Given a weighted connected graph $G=(V,E,w)$ with a set of $k$ terminals $T = \{t_1,t_2,\cdots,t_k\}\subset V$,
the Steiner Point Removal (SPR) problem seeks to construct a minor $G'=(T,E',w')$ of the graph $G$, such that for any $t_i,t_j\in T$,
$\dist_G(t_i,t_j) ~\le~ \dist_{G'}(t_i,t_j) ~\le~ \kappa \cdot \dist_G(t_i,t_j)$ with the minimum possible distortion $\kappa$.
Our main result is that for general graphs, the distortion is at most $\O(\log^2 k)$:

\begin{theorem}\label{thm:main}
For any weighted graph with $k$ terminals, the Steiner Point Removal problem can be solved with distortion at most $\O(\log^2 k)$.
More precisely, when $k$ is sufficiently large, the distortion is at most $(2\times 10^8)\cdot \log^2 k$.
\end{theorem}
We note that our analysis is rather loose on the constant factor,
so there should be plenty of room for reducing the constant $2\times 10^8$.
But we have not yet done so for a cleaner analysis.

\medskip

\parabold{Notations.}
By using a consistent tie-breaking rule (e.g., edge weight perturbation), we can assume that
there is a unique shortest path between any two vertices in $G$.
We let $\SP(v_i,v_j)$ denote the unique \emph{directed} shortest path in $G$ from vertex $v_i$ to vertex $v_j$.

For any $V'\subset V$, let $G[V']$ be the subgraph of $G$ induced by $V'$.
If $t\in V'$, let $B_{G[V']}(t,R) := \{v\in V' ~|~ \dist_{G[V']}(t,v)\le R\}$.

\medskip

\parabold{Terminal-centered Minors.}
As KKN showed, SPR problem is equivalent to finding the \emph{terminal-centered minor} of $V$
that minimizes the quantity $\max_{t_i,t_j\in T} \frac{\dist_{G'}(t_i,t_j)}{\dist_G(t_i,t_j)}$.
A terminal-centered minor of $G=(V,E,w)$ is specified by a \emph{terminal-centered partition} of $V$ into exactly $k$ sets,
which satisfies
(a) each set contains exactly one of the terminals, and thus the set containing $t_j$ can be naturally denoted by $V_j \equiv V_{t_j}$;
(b) for each $j\in [k]$, the induced graph $G[V_j]$ is connected.
Then the terminal centered minor $G'=(T,E',w')$ is formed
by contracting the vertices in each $V_j$ into a single vertex, which is identified with $t_j$.
For any $t_i,t_j\in T$, $(t_i,t_j)\in E'$ if and only if there exists an edge $(u,v)\in E$
such that one of its endpoints belongs to $V_i$ and the other endpoint belongs to $V_j$.
For each $(t_i,t_j)\in E'$, its weight $w'((t_i,t_j)) := \dist_G(t_i,t_j)$.

\medskip

\parabold{Exponential Random Variables.}
An exponential random variable (ERV) with mean $\mu$ is the probability distribution
with density function $f(x) = \mu^{-1}\cdot e^{-x/\mu}$ on $x\ge 0$, and $f(x)=0$ on $x<0$.
We denote the distribution by $\Exp(\mu)$.
ERV enjoys two important properties:
(a) closeness under scaling: the random variable $\Exp(\mu)$ follows the same distribution as the random variable $\mu\cdot \Exp(1)$, and
(b) memoryless property: for any $a,b\ge 0$, $\prob{~\Exp(\mu) > a + b ~|~ \Exp(\mu) > a~} = \prob{~\Exp(\mu) > b~}$.
We will also use the following inequality: for any $\kappa\ge 0$, $\prob{~\Exp(\mu)~\le~ \kappa\mu~} \le \kappa$.

\section{Algorithm and Technical Overview}

{\LinesNotNumbered
	\begin{algorithm}[H]
		\textbf{INPUT: }A weighted graph $G=(V,E,w)$, and a set of terminals $T=\{t_1,\cdots,t_k\}\subset V$.\\
		\textbf{OUTPUT: }A terminal-centered partition $\{V_1,\cdots,V_k\}$ of $V$.
		\begin{algorithmic}
			\State \nl Set $r~\la~ 1 + \delta / \log k$, where $\delta = 1/2$.
			\State \nl Set $D~\la~ \frac \delta{100\log k} \cdot \min_{v\in V\setminus T} D_v$.
			\State \nl For each $j\in [k]$, set $V_j~\la~\{t_j\}$, and set $R_j~\la~ 0$.~~~~\emph{//~$R_j \equiv R_{t_j}$.}
			\State \nl Set $V_\perp~\la~V\setminus\left(\cup_{j=1}^k V_j\right)$.~~~~\emph{//~$V_\perp$ keeps track of the set of unassigned vertices.}
			\State \nl Set $\ell ~\la~ 0$.
			\State \nl \While{$\left(\cup_{j=1}^k V_j\right) ~\neq~ V$}{
				\emph{//We call this round by Round-}$(D\cdot r^\ell)$\;
				\ForAll{$j\in [k]$}{
					\State ~~~~~~Choose independently at random $q^\ell_j~\sim~\Exp(D\cdot r^\ell)$.
					\State ~~~~~~Set $R_j~\la~R_j + q^\ell_j$.
					\State ~~~~~~Set $V_j~\la~V_j ~\cup~ B_{G[V_\perp \cup V_j]}(t_j,R_j)$.
					\State ~~~~~~Set $V_\perp~\la~V\setminus\left(\cup_{j=1}^k V_j\right)$.
				}
				Set $\ell~\la~\ell+1$.
			}
			\State \nl \Return $\{V_1,\cdots,V_k\}$.
		\end{algorithmic}
		\caption{The Ball-Growing Algorithm that produces a terminal-centered partition.}
		\label{alg:ball-growing-KKN}
	\end{algorithm}}

\subsection{Algorithm}

Our first step is to use an algorithm of Krauthgamer et al.~\cite[Theorem 2.1]{KNZ2014} to obtain a minor of $G$, denoted by $\Gs$,
such that all terminal distances are preserved exactly, and $\Gs$ contains only $\O(k^4)$ non-terminals
--- note that this number does not depend on $|V|$ and is in $\polyk$.
In the rest of this paper, we identify $\Gs$ as $G$.

Next, we use the randomized ball-growing algorithm of KKN
(with a few adjustments on parameters, see Algorithm \ref{alg:ball-growing-KKN})
on the graph $G$ to construct a terminal-centered partition.
In the course of the algorithm, a vertex $v$ is \emph{assigned} to terminal $j$ if $v\in V_j$;
we say a vertex $v$ is \emph{assigned} if it is assigned to some terminal, and we say $v$ is \emph{unassigned} otherwise.
As already noted in Algorithm \ref{alg:ball-growing-KKN},
we always refer to a round by Round-$\mu$, where $\mu$ is the mean of the ERV generated in that round.
We abuse notation a bit: for any positive real number $z$, by Round-$z$ we refers to the first Round-$\mu$ in the algorithm such that $\mu\ge z$;
note that $\mu\le \frac{3}{2}z$.

\subsection{Families of Bad Events}\label{sect:bad-events}

We define the following $\polyk$ ``bad events'' ($C_1>1$ and $C_2$ are constants to be determined):
\begin{enumerate}
\item[\textbf{(A)}] For each non-terminal $v$ in $G$, let its distance to the its nearest terminal be $D_v$,
and let $t$ be a terminal which is \emph{far} from $v$; precisely, $\dist_G(v,t) \ge C_1 \cdot D_v$.
Let $\calEfar_{v,t}$ denote the ``bad event'' that $v$ is assigned to $t$.
\item[\textbf{(B)}] For each non-terminal $v$ in $G$, let $\calEearly_v$ denote the ``bad event'' that
$v$ is assigned before the end of Round-$(C_2 D_v \delta/\log k)$.
\end{enumerate}

We will define another family of ``bad events'', but before doing so, we need the following setups,
which are modified from KKN.\footnote{Some of our terminologies are same as those used by KKN,
but they have slightly different definitions/meanings.}

For each pair of terminals $t_i,t_j$, let the vertices on $\SP(t_i,t_j)$, when listed in the direction from $t_i$ to $t_j$,
be $v_0=t_i,v_1,v_2,\cdots,v_{L-1},v_L=t_j$.
The vertices $v_1,\cdots,v_{L-1}$ are partitioned into sets, while each set contains consecutive vertices of $\SP(t_i,t_j)$.
For each such set $Q = \{v_a,v_{a+1},\cdots,v_b\}$,
we say its \emph{internal length} is the distance between $v_a$ and $v_b$,
while its \emph{external length} is the distance between $v_{a-1}$ and $v_{b+1}$.
(It is possible to have $v_a = v_b$; in this case, the internal length is $0$.)

We require each $Q$ in the partition satisfies the following requirement:
$Q$ contains a non-terminal $v$ such that the internal length of $Q$ is \underline{at most} $C_2 D_v \delta/(5\log k)$
but the external length of $Q$ is \underline{at least} $C_2 D_v \delta/(5\log k)$.
Such partition can be obtained by a simple greedy sweeping along the directed path $\SP(t_i,t_j)$,
where we always take $v\equiv v(Q)$ to be the first vertex of each set $Q$.
The algorithm that constructs $\Gs$ guarantees that there are at most $\O(k^2)$ vertices on $\SP(t_i,t_j)$,
and thus the cardinality of the partition is bounded by the same number. We denote the partition by $\calQ(t_i,t_j)$.

For each $Q = \{v_a,v_{a+1},\cdots,v_b\}\in \calQ(t_i,t_j)$, before the execution of the algorithm, all vertices in $Q$ are \emph{active}.
We will then iteratively determine how these vertices are turned to \emph{inactive}.
On the course of the algorithm, we say that a terminal $t$ \emph{reaches} $Q$ at a moment
when some \underline{active vertex} in $Q$ is assigned to $V_t$; when this happens, we let
$$
q_{\min} := \min~\left\{ q ~|~ v_q \text{ is active just before $t$ reaches $Q$, and it is assigned to $t$ after $t$ reaches $Q$} \right\};
$$
we define $q_{\max}$ analogously.
Then \underline{all} vertices $\{~v_q ~|~ q\in [q_{\min},q_{\max}]~\}$ are turned to inactive.
Also, we let $\TD(q_{\min},q_{\max}) := \SP(v_{q_{\min}},t) + \SP(t,v_{q_{\max}}) + (v_{q_{\max}},v_{q_{\max}+1})$
to be the \emph{terminal-detour} created by this $t$'s reaching on $Q$.
To avoid confusion, we make the following remark: when a vertex is assigned, it must be inactive,
but it is possible for a vertex to be inactive but remains unassigned.

Now we are ready to define the last family of ``bad events'': ($C_3$ is a constant to be determined)
\begin{enumerate}
\item[\textbf{(C)}] Let $\calEmany_Q$ denote the ``bad event'' that at least $C_3\log k$ distinct terminals reach $Q$.
\end{enumerate}

For the convenience of forthcoming discussion, whenever there are two \emph{consecutive} terminal-detours
$\TD(q_1,q_2)$ and $\TD(q_2+1,q_3)$ which involve the same terminal $t$, then the two terminal-detours
are merged into a new one, which is $\SP(v_{q_1},t) + \SP(t,v_{q_3}) + (v_{q_3},v_{q_3+1})$.
By repeating this merging process until no consecutive terminal-detours involve the same terminal,
we are sure that for any two consecutive terminal-detours $\TD(q_1,q_2)$ and $\TD(q_2+1,q_3)$,
the two vertices $v_{q_2},v_{q_2+1}$ are assigned to two different terminals,
and the two terminals are adjacent in the graph $G'$ (due to the edge $(v_{q_2},v_{q_2+1})\in E$).

\subsection{Key Lemmas and Proof of Theorem \ref{thm:main}}

By suitably choosing the constants $C_1,C_2,C_3$, we can show that the above $\polyk$ bad events
each occurs with probability at most $\O(1/k^\beta)$ for some sufficiently large $\beta$.
Then by a simple union bound, we can show that with probability $1-\O(1/k)$, all these bad events are avoided.
As we will show in the proof of Theorem \ref{thm:main}, the distortion of any terminal-centered minor that avoids all bad events
is at most $1 + \frac{40C_3(C_1+1)}{C_2} \log^2 k ~=~ \O(\log^2 k)$.

\begin{lemma}\label{lem:far-bound}
Set $C_1 = 5400$. Then $\prob{~\calEfar_{v,t}~} ~\le~ 3/k^6$.
\end{lemma}

\begin{lemma}\label{lem:too-early-bound}
Set $C_2 = 1/27$. Then $\prob{~\calEearly_v~} ~\le~ 2/k^5$.
\end{lemma}

\begin{lemma}\label{lem:too-many-bound}
Set $C_3 = 30$. Then $\prob{~\calEmany_P~|~\forall v,~\calEearly_v\text{\emph{ does not occur}}~} ~\le~ 1/k^5$.
\end{lemma}

The proofs of the above three lemmas will be provided in Section \ref{sect:analysis}.
We follow KKN's method closely to prove Lemma \ref{lem:too-many-bound}.
To prove Lamma \ref{lem:far-bound} and Lemma \ref{lem:too-early-bound},
we use the following two lemmas about the tail bounds of sum of independent ERV;
their proofs will be presented in Section \ref{sect:erlang}.
We note that sum of independent ERV with general \emph{distinct} means follows \emph{hypoexponential distribution},
for which complicated tail bounds are known, but none of them is simple enough for us to use directly.
Thus, we derive simpler tail bounds ourselves. This is done by a proper translation to finding tail bound of
sum of independent ERV with equal means, which is known to follow an \emph{Erlang distribution} and admits a simpler tail bound.

\begin{lemma}\label{lem:erlang-lower-tail}
For any positive integer $m$, let $E_1,E_2,E_3,\cdots,E_m$ be $m$ independent exponential random variables,
each has mean at least $A > 0$ (the means can be distinct). Then for any $\kappa \le 1/4$,
$$
\prob{~\sum_{i=1}^m E_i ~\le~ \kappa\cdot Am~} ~\le~ \frac{4}{3\sqrt{2\pi m}}\cdot (3\kappa)^m ~\le~ (3\kappa)^m.
$$
\end{lemma}

\begin{lemma}\label{lem:erlang-upper-tail}
Let $R_1,R_2,R_3,\cdots$ be a sequence of independent exponential random variables with means
$A,A/r,A/r^2,\cdots$, where $A>0$, and $r=1+\delta / \log k$ with $\delta \le 1/2$ and $k\ge 3$.
For any sufficiently large $k$, and for any constant $M \ge 18$,
$$
\prob{~\sum_{i=1}^\infty R_i ~\ge~ M\cdot \frac{A\log k}{\delta}~} ~\le~ \frac{2}{k^{M/(12\delta)+3}}.
$$
\end{lemma}

Note that in Lemma \ref{lem:erlang-upper-tail}, the expected value of
$\sum_{i=1}^\infty R_i$ is $\frac{Ar}{r-1} ~\le~ \frac{2A\log k}{\delta}$.
In other words, the lemma states that the probability for the sum
exceeding $M$ times of its expected value is at most $k^{-\Theta_\delta(M)}$.

The intuition behind proof of Lemma \ref{lem:far-bound} is:
when $t$ is \emph{far away} from $v$ but there are some other terminals \emph{near} $v$,
it looks highly likely that $v$ is assigned (to the one of the near terminals)
well before $t$ grows its ball radius large enough to come close to $v$.
We will verify this natural intuition rigorously.
Formally, we design a random variable to track how $v$ is being approached by the sets in the terminal-centered partition;
the random variable has initial value zero, and its value is at least $D_v$ when $v$ is assigned.
By showing that the random variable dominates some sum of independent ERV, we prove that with high probability,
$v$ is assigned early enough (using Lemma \ref{lem:erlang-lower-tail})
while $R_t$ is still below $C_1\cdot D_v$ (using Lemma \ref{lem:erlang-upper-tail}).

Lemma \ref{lem:too-early-bound} is almost a direct corollary of Lemma \ref{lem:erlang-upper-tail}.

\subsection{Proof of Theorem \ref{thm:main}}

We finish this technical overview by proving Theorem \ref{thm:main} using Lemmas \ref{lem:far-bound}---\ref{lem:too-many-bound}.

Let the terminals be $t_1,\cdots,t_k$ and let $V_1,\cdots,V_k$ be a terminal-centered partition which avoids all the above bad events
--- such parition exists for any sufficiently large $k$, since by Lemmas \ref{lem:far-bound}---\ref{lem:too-many-bound},
the probability that any of the bad events occur is at most
$$
\O(k^5)\cdot \frac{3}{k^6} ~+~ \O(k^4)\cdot \frac{2}{k^5} ~+~ \binom{k}{2}\cdot \O(k^2) \cdot \frac{1}{k^5}
~=~ \O\left( \frac{1}{k} \right) ~\ll~ 1.
$$

For each pair of terminals $t_i,t_j$, the length of $\SP(t_i,t_j)$ is bounded below
by half of the external lengths of all sets in $\calQ(t_i,t_j)$
(since some of the edges on $\SP(t_i,t_j)$ are double counted, we need the factor of half), i.e.,
$$
\dist_G(t_i,t_j) ~\ge~ \frac 12~\sum_{Q\in \calQ(t_i,t_j)} \frac{C_2 D_{v(Q)} \delta}{5\log k}.
$$

Recall that the vertices on $\SP(t_i,t_j)$, when listed in the direction from $t_i$ to $t_j$,
are\\
$v_0=t_i,v_1,v_2,\cdots,v_{L-1},v_L=t_j$.
For each $Q = \{v_a,v_{a+1},\cdots,v_b\}\in \calQ(t_i,t_j)$, note that conditioning on the avoidance of $\calEmany_Q$,
there is a directed path from $v_a$ to $v_{b+1}$ in $G$ which is composed of at most $C_3 \log k$ terminal-detours;
we denote this directed path by $P_Q$.

Concatenating $(t_i,v_1)$ with all $P_Q$, where $Q$ runs over $\calQ(t_i,t_j)$,
forms a directed path from $t_i$ to $t_j$ in $G$, denoted by $\detour(t_i,t_j)$.\footnote{Note that
this directed path may not be \emph{simple}, i.e., it may traverse a terminal or an edge for multiple times.
This does not matter for us, since the purpose for introducing this path is to give an over-estimate of $\dist_{G'}(t_i,t_j)$.}
Due to the terminal-detour-merging process described at the end of Section \ref{sect:bad-events},
every two consecutive terminal-detours involve two different terminals, and the two terminals are adjacent in $G'$.
Thus $\detour(t_i,t_j)$, when contracted according to the underlying terminal-center partition, forms a path from $t_i$ to $t_j$ in $G'$.
This path in $G'$ has length bounded above by the length of $\detour(t_i,t_j)$ in $G$, due to the definition of $w'$ in $G'=(T,E',w')$.
Conditioned on the avoidances of all the bad events, the length of $\detour(t_i,t_j)$ is upper bounded by
$$
\dist_G(t_i,t_j) ~+~ \sum_{Q\in\calQ(t_i,t_j)}\left[2(C_1+1)\cdot D_{v(Q)}\right]\cdot (C_3\log k),
$$
and hence the distortion is bounded above by
$$
1 + \frac{\sum_{Q\in\calQ(t_i,t_j)}\left[2(C_1+1)\cdot D_{v(Q)}\right]\cdot
(C_3\log k)}{\frac 12~\sum_{Q\in \calQ(t_i,t_j)} \frac{C_2 D_{v(Q)} \delta}{5\log k}}
~\le~ 1 + \frac{40C_3(C_1+1)}{C_2} \log^2 k,
$$
which is at most $(2\times 10^8)\cdot \log^2 k$.

\section{Analysis}\label{sect:analysis}

\newcommand{\ts}{t_*}
\newcommand{\us}{u_*}
\newcommand{\ds}{d_*}

\subsection{Proof of Lemma \ref{lem:far-bound}}

\parabold{Setup.}
Let $\ts$ denote the nearest terminal to $v$ in graph $G$,
and let $P$ denote the \emph{directed} shortest path from $\ts$ to $v$.
By definition, the length of $P$ is $D_v$.
We use the variable $\us$ to track the vertex in $P$ which is assigned and furthest away from $\ts$.
At the beginning, the only assigned vertex in $P$ is $\ts$, thus $\us=\ts$;
when $\us = v$, $v$ is assigned.

Let $t_q$ be the variable tracking the terminal which $\us$ is assigned to.
Let $d_q$ denote the distance between $t_q$ and $\us$ in $G[V_q]$, then consider the following random variable:
$$
\ds ~:=~ \dist_G(\ts,\us) ~+~ R_q ~-~ d_q.
$$
Note that $\ds$ never decreases, and $\ds \ge D_v$ if and only if $v$ is assigned.

\medskip

\parabold{Step 1.} In this step, we show that after Round-$T$, where $T = 200 D_v \delta / \log k$,
$\ds \ge D_v$ occurs with probability at least $1-1/k^6$.

There are two scenarios in which $\ds$ increases: (a) when $R_q$ increases, while $\us$ might remain unchanged or not;
or (b) when some vertex on $P$ which is further away from $\ts$ than $\us$ is assigned to some terminal $t_{q'}$, where $q'\neq q$.

In scenario (a), $\ds$ increases by exactly the amount of the ERV that raises $R_q$.

In scenario (b), since $\us$ is changed, to avoid confusion, we use $\us^{\text{old}}$ to denote the value of $\us$ before the change.
Let $u'$ denote the vertex on the directed path $\SP(\us^{\text{old}},v)$
which is newly assigned to $t_{q'}$ and nearest to $t_{q'}$ in $G[V_{q'}]$.
Due to the memoryless property of ERV, the increment of $\ds$ is the distance between $\us^{\text{old}}$ and $u'$, 
plus a random variable that follows an exponential distribution with the same mean as the ERV that raises $R_{q'}$.

On the other hand, observe that in each round, at least one of the two scenarios occurs
(and for (b), it might occur multiple times in one round).
Combined with the last paragraph, after Round-$T$, the value of $\ds$ is lower bounded by the sum of independent ERV of the form
$\sum_{i=1}^m \Exp(T/r^{i-1})$, where $m = \ceil{(\log k)/\delta}$.
Note that the mean of $T/r^{m-1} \ge T/3 = 200 D_v \delta / (3\log k)$.
Thus, by Lemma \ref{lem:erlang-lower-tail},
$$
\prob{~\ds < D_v~} ~\le~ \prob{~\sum_{i=1}^m \Exp(T/r^{i-1}) ~<~ \frac{3}{200} \cdot \frac{200 D_v \delta}{3\log k}\cdot m ~}
~\le~ \left( \frac{9}{200} \right)^{\log k / \delta} ~\le~ \frac{1}{k^6}.
$$

\parabold{Step 2.} Conditioned on the highly probable event analyzed in Step 1,
for the event $\calEfar_{v,t}$ to occur, a necessary (but not sufficient) condition is:
the radius $R_t$ grows beyond $C_1 D_v = 5400 D_v$ by the end of Round-$T$.
By Lemma \ref{lem:erlang-upper-tail}, for sufficiently large $k$, this occurs with probability at most
$$
\prob{~\sum_{i=0}^\infty \Exp\left(\frac{3T}{2}\left/r^i\right.\right) ~\ge~ 18\cdot \frac{300 D_v \delta}{\log k} \cdot \frac{\log k}{\delta}~}
~\le~ \frac{2}{k^6}.
$$

Thus, we can conclude that $\prob{~\calEfar_{v,t}~}~\le~ 1/k^6 + 2/k^6 ~=~ 3/k^6$.

\subsection{Proof of Lemma \ref{lem:too-early-bound}}

Lemma \ref{lem:too-early-bound} can be proved almost directly from Lemma \ref{lem:erlang-upper-tail}.
For the event $\calEearly_v$ to occur, a necessary (but not sufficient) condition is:
there exists a terminal $t$ with radius $R_t$ grows beyond $D_v$ by the end of Round-$(C_2 D_v \delta/\log k)$.
By Lemma \ref{lem:erlang-upper-tail}, this occurs with probability at most
\begin{align*}
&\sum_{t\in T}~\prob{~\sum_{i=0}^\infty \Exp\left(\frac{3C_2 D_v\delta}{2\log k}\left/r^i\right.\right) ~\ge~ D_v~}\\
&~~~~~~~~=~ k ~\cdot~ \prob{~\sum_{i=0}^\infty \Exp\left(\frac{3C_2 D_v\delta}{2\log k}\left/r^i\right.\right) ~\ge~
\frac{2}{3C_2}\cdot \frac{3 C_2 D_v \delta}{2 \log k} \cdot \frac{\log k}{\delta}~} ~~\le~~ k ~\cdot~ \frac{2}{k^6} ~=~ \frac{2}{k^5}.
\end{align*}

\subsection{Proof of Lemma \ref{lem:too-many-bound}}

The idea behind the proof of Lemma \ref{lem:too-many-bound} follows closely from~\cite[Sections 3.1.1 and 3.1.2]{KKN2015},
but there are some small differences in the details. For completeness, we provide a self-contained proof.

Let $Q = \{v_a,v_{a+1},\cdots,v_b\}$.
On the course of the algorithm, a \emph{maximal active subset} of $Q$ is a subset $Q' = \{v_q,v_{q+1},\cdots,v_r\}\subset Q$
such that (a) all vertices in $Q'$ are active; (b) $v_{q-1}$ is inactive or $q = a$; and
(c) $v_{r+1}$ is inactive or $r = b$.
We let $Z_Q$ count the number of maximal active subsets of $Q$ on the course of the algorithm.
At the beginning, the only maximal active subset of $Q$ is $Q$ itself, so $Z_Q = 1$.
Whenever $Z_Q = 0$, we obtain a collection of terminal-detours which are later used to form $P_Q$,
which we have defined in the proof of Theorem \ref{thm:main}.

For each $\ell\in\nn$, let $t^\ell$ denote the $\ell$-th terminal that reaches $Q$,
then let $Y_\ell$ denote the indicator random variable that upon the reach of $t^\ell$ on $Q$, $Z_Q$ strictly decreases.
We note that when $Y_\ell = 0$, by inspecting the process of how vertices are turned inactive (see Section \ref{sect:bad-events}),
it is not hard to see that $Z_Q$ increases by at most $1$,
and furthermore, it must be the case that \emph{some but not all} vertices in one maximal active subset get assigned to $t^\ell$.\footnote{An
elaborative explaination: (1) If $v_{q_{\min}}$ and $v_{q_{\max}}$ belong two different maximal active subsets (denoted as $S_1,S_2$),
then all maximal active subsets strictly between $S_1,S_2$ are turned to inactive,
while $S_1,S_2$ might shrink to a smaller maximal active subset or be turned to fully inactive.
When either $S_1,S_2$ is turned to fully inactive, $Z_Q$ must drop by at least $1$. Thus, whenever $Z_Q$ does not strictly decrease,
some vertices in either $S_1,S_2$ remain unassigned.
(2) If $v_{q_{\min}}$ and $v_{q_{\max}}$ belong to the same maximal active subset $S_1$, then it's clear that $Z_Q$ can increase by at most $1$,
and when $Z_Q$ strictly increases, some vertices in $S_1$ remain unassigned.
}

Note that for any vertex $v\in Q$, $D_v ~\ge~ D_{v(Q)} \cdot (1-C_2 \delta / 5\log k) ~\ge~ 99 D_{v(Q)} / 100$.
Conditioned on the avoidances of all $\calEearly_v$, a vertex in $Q$ can be assigned only after Round-$(99 C_2 D_{v(Q)}\delta/(100\log k))$.
Also, note that the internal length of each maximal active subset is at most the internal length of $Q$,
which is at most $C_2 D_{v(Q)}\delta/(5\log k)$. Hence, by the memoryless property of ERV, we have
$$
\prob{~Y_\ell=0~} ~\le~ \prob{~\Exp\left(\frac{99 C_2 D_{u(Q)}\delta}{100 \log k}\right)
~\le~ \frac{C_2 D_{u(Q)}\delta}{5\log k} ~} ~\le~ \frac{20}{99}.
$$

\newcommand{\Ybar}{\overline{Y}}

Next, for an integer $N$ to be determined, let $\Ybar_N := \sum_{\ell=1}^{N} Y_\ell$.
Note that when $\Ybar_N > N/2$, then $Z_Q = 0$.
Thus, $\prob{~Z_Q\ge 1\text{ after $Q$ being reached for $N$ times}~} ~\le~\prob{~\Ybar_N \le N/2~}$,
which is bounded above by $\prob{~\mathsf{Binom}(N,79/99)~\le~ N/2~}$.\footnote{$\mathsf{Binom}(N,p)$ is the Binomial distribution
followed by $\sum_{\ell=1}^{N} X_\ell$, where each $X_\ell$ is an independent random variable that takes value $1$ with probability $p$,
and it takes value $0$ otherwise.}
 By the Hoeffding's inequality,
$$
\prob{~\mathsf{Binom}\left(N,\frac{79}{99}\right)~\le~ \frac{N}{2}~} ~\le~ e^{-2(79/99 - 1/2)^2 N}.
$$
By choosing $N = 30\log k$, the RHS of the above inequality is at most $1/k^5$.

\section{Tail Bounds of Sum of Exponential Random Variables}\label{sect:erlang}

In this section, we provide proofs of Lemma \ref{lem:erlang-lower-tail} and Lemma \ref{lem:erlang-upper-tail}.

\begin{pfof}{Lemma \ref{lem:erlang-lower-tail}}
First, since ERV enjoys the property of closeness under scaling, it suffices to prove the lemma for $A=1$.

It is well-known that sum of independent ERV with \emph{equal} means follows an Erlang distribution. Thus,
$$
\prob{~\sum_{i=1}^m E_i ~\le~ \kappa m~} ~\le~ \prob{~\sum_{i=1}^m \Exp(1) ~\le~ \kappa m~}
~=~ \prob{~\erlang(m,1) ~\le~ \kappa m~} ~=~ \frac{\gamma(m,\kappa m)}{(m-1)!},
$$
where $\gamma(\cdot,\cdot)$ is the \emph{lower incomplete Gamma function},
which admits a power series expansion given below~\cite{incomplete-Gamma}.
Together with the assumption $\kappa \le 1/4$, it yields
\begin{align*}
\frac{\gamma(m,\kappa m)}{(m-1)!}
&~=~ \frac{e^{-\kappa m} (\kappa m)^m}{(m-1)!} \cdot \sum_{\ell=0}^{\infty} \frac{(\kappa m)^\ell}{m(m+1)\cdots(m+\ell)}\\
&~\le~ \frac{e^{-\kappa m} (\kappa m)^m}{m!}\cdot \left[ 1 + \frac 14 + \frac 1{16} + \cdots\right]
~=~ \frac{4 e^{-\kappa m} (\kappa m)^m}{3\cdot m!}.
\end{align*}
By the Stirling's approximation, $m!\ge \sqrt{2\pi m}\cdot (m/e)^m$, thus we can complete the proof:
$$
\frac{4 e^{-\kappa m} (\kappa m)^m}{3\cdot m!} ~\le~ \frac{4}{3\sqrt{2\pi m}}\cdot [e^{1-\kappa}\kappa]^m
~\le~ \frac{4}{3\sqrt{2\pi m}}\cdot (3\kappa)^m.
$$
\end{pfof}

To prove Lemma \ref{lem:erlang-upper-tail}, we need the following intermediate lemma.

\begin{lemma}\label{lem:erlang-upper-tail-inter}
Let $E_1,E_2,E_3,\cdots,E_m$ be $m$ independent exponential random variables,
each has mean at most $A > 0$.
Then for any sufficiently large $m$, and for any $C\ge 6$,
$$
\prob{~\sum_{i=1}^m E_i ~\ge~ C\cdot Am~} ~\le~ \frac{2}{(C-1)\cdot \sqrt{2\pi m}}\cdot \frac{1}{e^{Cm/2}} ~\le~ \frac{1}{e^{Cm/2}}.
$$
\end{lemma}
\begin{pf}
Again, it suffices to prove the lemma for $A=1$.
Note that
$$
\prob{~\sum_{i=1}^m E_i ~\ge~ Cm~} ~\le~ \prob{~\sum_{i=1}^m \Exp(1) ~\ge~ Cm~}
~=~ \prob{~\erlang(m,1) ~\ge~ Cm~} ~=~ \frac{\Gamma(m,Cm)}{(m-1)!},
$$
where $\Gamma(\cdot,\cdot)$ is the \emph{upper incomplete Gamma function}.
Tricomi~\cite{Tricomi1950} (see also~\cite{Gautschi1998}) showed that for any constant $C \ge 6$,
$$
\Gamma(m,Cm) ~=~ \frac{e^{-Cm}~(Cm)^m}{Cm-m+1} \cdot (1+o_m(1)).
$$
By the Stirling's approximation, $m!\ge \sqrt{2\pi m}\cdot (m/e)^m$. Thus,
for any sufficiently large $m$,
$$
\frac{\Gamma(m,Cm)}{(m-1)!} ~\le~ \frac{2 e^{-Cm}(Cm)^m}{(C-1)\cdot m!}
~\le~ \frac{2}{(C-1)\cdot \sqrt{2\pi m}}\cdot [e^{1-C}C]^m.
$$
When $C\ge 6$, $e^{1-C}C \le e^{-C/2}$, and we are done.
\end{pf}

\begin{pfof}{Lemma \ref{lem:erlang-upper-tail}}
Again, it suffices to prove the lemma for $A=1$.
Let $m := \floor{(\log k) / \delta} ~\ge~ \frac{\log k}{2\delta}$.
For $j\in\mathbb{N}$, let
$$
\calE_j \text{ denote the following event: }\sum_{r=(j-1)m+1}^{jm} R_r
~\ge~ \frac{1}{2^{j-1}} \cdot \left( \frac{1}{2} + \frac{j}{2} \right) \cdot \frac{M\log k}{3\delta}.
$$
Due to the equality $\sum_{j=1}^\infty (1+j)/2^j ~=~ 3$, if $\sum_{i=1}^\infty R_i ~\ge~ \frac{M \log k}{\delta}$,
then at least one of $\calE_1,\calE_2,\cdots$ must hold.
Hence,
$$\prob{~\sum_{i=1}^\infty R_i ~\ge~ \frac{M \log k}{\delta}~} ~\le~ \sum_{j=1}^\infty~\prob{~\calE_j~}.$$
Next, observe that $\expect{R_{(j-1)m+1}} ~\le~ (1+1/m)^{-(j-1)m} ~\le~ 1/2^{j-1}$. Hence, by Lemma \ref{lem:erlang-upper-tail-inter},
$$
\prob{~\calE_j~}
~\le~ 1\left/\exp{\left[\left( \frac{1}{2} + \frac{j}{2} \right) \cdot \frac{M\log k}{3\delta} \cdot \frac{1}{2}\right]}\right.
~=~ k^{-(1+j)M/(12\delta)} ~\le~ \frac{1}{k^{M/(12\delta)}} \cdot \frac{1}{k^{3j}},
$$
and thus
$$\prob{~\sum_{i=1}^\infty R_i ~\ge~ \frac{M \log k}{\delta}~} ~\le~ \frac{1}{k^{M/(12\delta)+3}}\left(1 + \frac{1}{k^{3}} + \frac{1}{k^6} + \cdots\right) ~\le~ \frac{2}{k^{M/(12\delta)+3}}.$$
\end{pfof}

\section{Discussion}

If we are allowed to use only one brief sentence to summarize how the $\O(\log^2 k)$ distortion is arrived, it is:
the vertices in a short segment of length $\ell = \Theta(D_v/\log k)$ might be assigned to at most $\Theta(\log k)$ distinct terminals,
each is of distance of at most $\Theta(\ell \log k) = \Theta(D_v)$ away from the segment.

In our (failed) attempts to shave further $\log$,
one idea is to grow the ball radius more aggressively by setting $r = 1 + \Theta(1)$ instead of $r = 1 + \Theta(\log^{-1} k)$.
To motivate this idea, see in the proof of Lemma \ref{lem:far-bound} and by Lemma \ref{lem:too-early-bound},
a vertex $v$ is assigned in Round-$\left[\Theta(D_v / \log k)\right]$. This forced us to
set $\ell$ to be of the same magnitude for establishing Lemma \ref{lem:too-many-bound}.
By setting $r = 1 + \Theta(1)$, if the sum of ERV $\sum_{i=1}^\infty \Exp(\Theta(D_v)/r^i)$ were well concentrated around its mean,
we could hope that $v$ is assigned in Round $\Theta(D_v)$ and then we could set $\ell = \Theta(D_v)$.
Unfortunately, such sum of ERV is actually \emph{not} sufficiently well concentrated.

Another idea is to replace the exponential random variables, which have unbounded support,
with some other random variables, say those with bounded supports.
Then Chernoff-like tail bounds might be plausible. However, memoryless property is lost.
While I have my personal intuition that this approach might provide better results,
the loss of memoryless property makes devising a rigorous proof seem technically challenging.

This draws a parallel with the early development of \emph{queueing theory}.
Its fundamental results mostly apply to models in which the involved random variables are ERV ---
the reason is not because ERV is the most appropriate probability distribution that captures the reality,
but because the memoryless property of ERV makes analyses more admissible.
After more than a century of development, rich analytical insights and tools have been developed to handle more general queueing models;
it might be interesting to investigate whether such insights and tools can help us to ``grow balls for the better''.

\bibliographystyle{alpha}
\bibliography{bibitem}
\end{document}